\documentclass[12pt]{iopart}

\usepackage{iopams} 
\usepackage{graphicx,cite}
\usepackage{psfrag}
\usepackage{pstool}
\usepackage{subfigure}
\usepackage{color}
\newcommand{\ket}[1]{\left|#1\right\rangle}
\newcommand{\bra}[1]{\left\langle#1\right|}
\newcommand{\be}{\begin{equation}}
\newcommand{\ee}{\end{equation}}
\newcommand{\bea}{\begin{eqnarray}}
\newcommand{\eea}{\end{eqnarray}}
\newcommand{\lr}{\left(}
\newcommand{\rr}{\right)}
\newcommand{\ls}{\left[}
\newcommand{\rs}{\right]}
\newcommand{\half}{\frac{1}{2}}
\usepackage{tikz}
\usetikzlibrary{arrows}
\usetikzlibrary{decorations.pathmorphing}
\usepackage[hang,small,bf]{caption}
\begin{document}

\title{Scattering of coherent states on a single artificial atom}

\author{B. Peropadre$^{1}\footnote[3]{These two authors contributed equally to the manuscript.},$ J. Lindkvist$^{2}\S$, I.-C. Hoi$^2$, C. M. Wilson$^2$,  J. J. Garcia-Ripoll$^1$, P.Delsing$^2$ and G. Johansson$^2$}
\address{$^1$ Instituto de F{\'\i}sica Fundamental, CSIC, Calle Serrano 113-bis, Madrid
E-28006, Spain}
\address{$^2$ Department of Microtechnology and Nanoscience, Chalmers University of Technology, G\"oteborg}
\ead{Goran.L.Johansson@chalmers.se}
\begin{abstract}
In this work we theoretically analyze a circuit QED design where
propagating quantum microwaves interact with a single artificial atom,
a single Cooper pair box. In particular, we derive a master equation
in the so-called transmon regime, including coherent drives. 
Inspired by recent experiments, we then apply the master equation to describe the dynamics 
in both a two-level and a three-level approximation of the atom. In the two-level case, we also discuss how to measure photon antibunching in the reflected field and how it is affected by finite temperature and finite detection bandwidth. 
\end{abstract}
\date{\today}

\section{Introduction}
In recent years, the field of circuit quantum
electrodynamics\cite{clarke08,schoelkopf08} (circuit QED) has become
one of the most promising platforms in the study of light-matter interaction.
One of the most important breakthroughs in this field was the achievement of strong coupling between light and matter, or microwave photons and Josephson-based artificial atoms \cite{blais04,
  wallraff04}.  Since then, many experiments have been carried out within the framework of superconducting
circuits\cite{schuster05,schuster07,majer07,houck07,fink08},
revealing a wide variety of novel quantum
phenomena. Most of these experiments share a common feature, 
namely the interaction between artificial atoms and isolated
modes of the electromagnetic field in a cavity. 
Within circuit QED, there is now a growing interest in studying propagating
fields interacting with artificial atoms, due to \textit{e.g.} its potential
interest in Condensed Matter~\cite{LeHur11} and
all-optical quantum information ~\cite{knill01}. Theoretically,
coherent coupling between an atom or superconducting qubit and a one dimensional
continuum of modes has been discussed since some time ago\cite{shen05, shen05b, PhysRevA.73.043613, PhysRevA.82.063816, PhysRevA.85.043832, PhysRevA.85.023817}, and there exists now a growing number of experiments investigating this system in a circuit QED setup\cite{Astafiev10, AstafievPRL2010, AbdumalikovPRL2010, AbdumalikovPRL2011,hoi11, hoi12, Hoi122, Hoi123}. 

In this manuscript we report on an in-depth microscopic description of the coherent
coupling between a field propagating through an 
open transmission line and a superconducting artificial atom based on the single Cooper-pair box (SCB) \cite{bouchiat98,nakamura99,makhlin07,lehnert03,gunnarsson05,But87,Duty04}. In more detail, we analyze the so-called transmon regime \cite{koch07,houck07} and study the photon transport
properties of this system according to different approximations. On
the one hand, in
the two-level approximation, and under certain conditions, the qubit
behaves as a saturable mirror \cite{shen05, shen05b}. On the other hand,
including a second excited state of the transmon, we can effectively 
make the medium transparent for the incident
photons using a coherent control field on resonance with this second transition.
Finally, we also discuss how the the photon antibunching observed in the reflected 
field is reduced by finite temperature and finite detection bandwidth.
Our theoretical predictions are in full agreement with recent experiments~\cite{hoi11,hoi12,Hoi123}.  
 
This paper is organized as follows: In section~\ref{sec:model}, we derive the master equation of a SCB coupled to an open transmission line (TL). In section \ref{subsec:discrete_model}, we start from a discretized lumped-element description of the TL 
and in section~\ref{subsec:continuum_model} we proceed to the continuum description.
In section~\ref{subsec:voltage_biased_SCB}, we discuss the regime where the system can be described as a SCB weakly coupled to the the voltage of the TL, at the coupling point. We thus arrive at the Hamiltonian of a voltage biased SCB, weakly coupled to a bath of harmonic oscillators, \textit{i.e.} the electromagnetic modes of the TL. Making standard weak coupling approximations, we then derive a master equation on Lindblad form in section~\ref{subsection:MasterEquation} and attach a coherent drive in section~\ref{sec:coherentdrive}.
For simplicity, we go through these derivations considering a SCB at the end of a semi-infinite TL, but in section~\ref{sec:morelines} we discuss how the master equation can be straightforwardly extended to an arbitrary number of semi-inifinte TLs, all meeting at the SCB. In particular, this includes the important case of a single infinite TL.

In section~\ref{sec:applications}, we then apply the master equation to a few experimentally relevant cases\cite{hoi11,hoi12}. Section \ref{sec:SCB} is devoted to the reflection and transmission of a single near resonant coherent drive, while section~\ref{subsec:three-level} includes two coherent drives, where one is used to control the transmission of the other.
Finally in section~\ref{sec:second-order}, we investigate how the photon antibunching observed in the reflected field is influenced by finite temperature and finite detection bandwidth.

\section{Model}
\label{sec:model}
\begin{figure}[!ht]
  \begin{center}
 \includegraphics[width=0.7 \columnwidth]{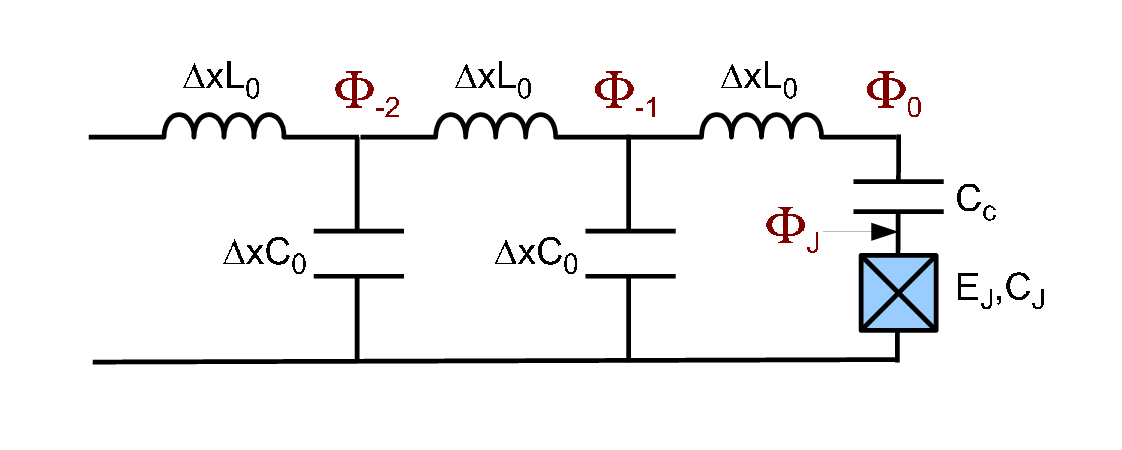}
\caption{Discretized circuit describing the interaction
  of a single Cooper-pair box with microwaves
  photons propagating in a semi-infinite transmission line.} 
\label{fig:circuit}
\end{center}
\end{figure}
In this section we present a general formalism of the light-matter
scattering in a one-dimensional continuum from a microscopic point of
view. We start from a Hamiltonian description, arriving at the
well-known input-output relations for the microwave field. We then
follow the usual approach \cite{gardinerzoller} to describe the joint state of the
light-matter system by introducing dissipation, resulting in the
standard quantum optical master equation.

\subsection{Discrete Circuit Model, Hamiltonian and Equations of motion}
\label{subsec:discrete_model}
 Consider a semi-infinite transmission line with characteristic inductance $L_0$ and capacitance $C_0$ per unit length.
 We discretize the transmission line\cite{YurkeDenker} in units of the small length  $\Delta x$, which we take to zero at the end of the calculation.  The transmission line nodes are numbered with negative integers, while the SCB island node has index $J$ and its Josephson junction has a capacitance $C_J$ to ground and a Josephson energy $E_J$. The SCB is coupled to the transmission line at the zeroth node, through the capacitance $C_c$, as depicted in Figure~\ref{fig:circuit}.

To describe the circuit dynamics we use the node fluxes $\Phi_\alpha(t)=\int^t dt' V_\alpha(t')$ as coordinates\cite{DevoretLesHouches1995}. They are the time integrals of the node voltages and although being less intuitive than the voltages, this choice greatly simplifies the description of the Josephson junction. Starting from a circuit Lagrangian we can derive the discrete circuit Hamiltonian\cite{JohanssonTornbergShumeikoWendin2006}
\be
H_{d}=\frac{(p_0+p_J)^2}{2C_J}+ \frac{p_0^2}{2C_c} -
E_J\cos{\left(\frac{2e}{\hbar}\Phi_J\right)}+\frac{1}{\Delta x} \sum_{n< 0} \frac{
    p_{n}^2}{2C_0} + \frac{(\Phi_{n+1} -\Phi_{n} )^2}{2L_0} ,
    \label{DiscreteHamiltonian}
\ee
where the charges $p_\alpha$ are the conjugate momenta to the node fluxes $\Phi_\alpha$, fulfilling the canonical commutation relations
\[ [\Phi_\alpha,p_\beta]=i\hbar \delta_{\alpha,\beta}, \ \  [\Phi_\alpha,\Phi_\beta]= [p_\alpha,p_\beta]=0, \]
where $ \delta_{\alpha,\beta}$ denotes Kronecker's delta. From the Hamiltonian we get Heisenberg's equations of motion for the transmission line operators ($n<0$)
\be
\partial_t \Phi_n = \frac{p_n}{\Delta x C_0}, \ \partial_t p_n = \frac{\Phi_{n-1}-2\Phi_n+\Phi_{n+1}}{\Delta x L_0},
\ee
and for the SCB operators
\begin{eqnarray}
\partial_t p_0&=&\frac{\Phi_{-1}-\Phi_0}{\Delta x L_0}, \label{dp0eq_v0} \\
\partial_t\Phi_0&=&\frac{p_0+p_J}{C_J}+\frac{p_0}{C_c}=\frac{C_\Sigma}{C_cC_J}p_0+\frac{p_J}{C_J}, \label{dphi0eq_v0}\\
\partial_t p_J&=&-E_J \frac{2e}{\hbar}\sin{\left(\frac{2e}{\hbar}\Phi_J\right)},  \label{dpjeq_v0} \\
\partial_t \Phi_J &=& \frac{p_0+p_J}{C_J},  \label{dphiJeq_v0}
\end{eqnarray}
where $C_\Sigma=C_c+C_J$.

\subsection{Continuum limit}
\label{subsec:continuum_model}
In the continuum limit $\Delta x \rightarrow 0$, the charge of each transmission line node will go zero together with the node capacitance. Thus we define a charge density field $p(x_n,t)=p_n(t)/\Delta x$ and a flux field $\Phi(x_n,t)=\Phi_n(t)$,
where we define the spatial coordinate $x_n=n\Delta x$ for $n<0$, along the transmission line. The continuum equations of motion for the transmission line ($x<0$) are
\begin{equation}
\partial_t p(x) = \frac{ \partial^2_{x} \Phi(x)}{L_0} , \ \ \partial_t\Phi(x)=\frac{p(x)}{C_0}.
\end{equation}
These are the equations of motion for the massless Klein-Gordon field, having freely propagating left and right moving solutions with velocity $v=1/\sqrt{L_0C_0}$. Therefore, we can write the general solution for $x<0$ as a linear combination of right and left moving second-quantized fields,
\begin{eqnarray}
\label{eq:fourierexpansion}
\Phi_{\rightleftarrows   }(x,t) &=& \sqrt{\frac{\hbar Z_0}{4\pi}}\int_{0}^\infty
\frac{d\omega}{\sqrt{\omega}}\left( a^{\rightleftarrows}_\omega e^{-i(\omega t \mp k_\omega x)} + h.c. \right) ,\nonumber\\
p_{\rightleftarrows   }(x,t) &=& -i \sqrt{\frac{\hbar Z_0}{4\pi}}\int_{0}^\infty
d\omega \sqrt{\omega}\left( a^{\rightleftarrows}_\omega e^{-i(\omega t \mp k_\omega x)} - h.c. \right) ,\nonumber\\
\end{eqnarray}                                                                                                                                                        
where $k_\omega=\omega/v$ and $Z_0=\sqrt{L_0/C_0}$ is the characteristic impedance of the
transmission line. The operators $a_{\omega}^{\leftrightarrows}$ annihilate a left/right-moving photon with frequency $\omega$, and  obey the bosonic canonical commutation relations, $[a^\leftarrow_\omega,(a_{\omega'}^\leftarrow)^{ \dagger}] = [a^\rightarrow_\omega,(a_{\omega'}^\rightarrow)^{ \dagger}]=\delta(\omega-\omega')$ and $[a^\leftarrow_\omega,(a_{\omega'}^\rightarrow)^{ \dagger}] = [a^\rightleftarrows_\omega,a_{\omega'}^\rightleftarrows]=0$. Finally, we note that in the continuum limit (\ref{dp0eq_v0}) changes into
\begin{equation}
\label{dp0eq_cont}
\partial_t p_0=-\frac{\partial_x \Phi(0^-)}{L_0}.
\end{equation}

To describe the system dynamics, we first need to specify the incoming, right moving field
\[\Phi^{in}(t)=\Phi_{\rightarrow}(0^-,t).\] 
Given this initial condition, we can then calculate the SCB dynamics, as well as the outgoing field
\[\Phi^{out}(t)=\Phi_{\leftarrow}(0^-,t),\] 
propagating to the left in the line.
The flux at $x=0$ is simply the sum of the incoming and outgoing flux fields
\begin{equation}
\label{FinalPhi0Eq}
\Phi_0(t)=\Phi(0^-,t)=\Phi^{in}(t)+\Phi^{out}(t)+V_{DC}t,
\end{equation}
where we for simplicity also explicitly extracted the dc voltage bias $V_{DC}$, implying that $\Phi^{in}$ and $\Phi^{out}$ have no dc components.
Now, solving for $p_0$ from (\ref{dphi0eq_v0}) gives
\begin{equation}
p_0=\frac{C_c C_J}{C_\Sigma} \left[ V_{DC}+\partial_t  \left(\Phi^{in}+\Phi^{out}\right)\right] -\frac{C_c}{C_\Sigma}p_J,
\label{dphi0eq_cont}
\end{equation}
and inserting this expression into (\ref{dphiJeq_v0}), we arrive at
\begin{equation}
\partial_t \Phi_J = \frac{p_J+C_c\left[V_{DC}+\partial_t  \left(\Phi^{in}+\Phi^{out}\right) \right]}{C_\Sigma}.
\label{fulldPhiJdt}
\end{equation}
We then insert $\Phi_0$ from (\ref{FinalPhi0Eq}) in (\ref{dp0eq_cont}) and arrive at
\begin{equation}
\partial_t p_0 =-\frac{\partial_x \Phi(0^-)}{L_0}=\frac{\partial_t \left(\Phi^{in}-\Phi^{out}\right)}{Z_0}, \label{dp0eq_cont}
\end{equation}
where we used the relation $\partial_x\Phi_\rightleftarrows(0^-)=\mp v^{-1}\partial_t\Phi_\rightleftarrows(0^-)$, to change the spatial derivative into a time derivative. 
Inserting the expression for $p_0$ from (\ref{dphi0eq_cont}) into the left hand side of this equation and integrating once with respect to time leads to
\begin{equation}
\Phi^{out}=\Phi^{in}+Z_0 \frac{C_c}{C_\Sigma}p_J-\tau_{RC} \partial_t  \left(\Phi^{in}+\Phi^{out}\right),
\label{fullPhiOutEq}
\end{equation}
where the time $\tau_{RC}=C_c C_J Z_0/C_\Sigma$ is the characteristic RC-time for discharging the SCB through the transmission line.
Equations (\ref{dpjeq_v0}), (\ref{fulldPhiJdt}) and (\ref{fullPhiOutEq}) in principle give the full time evolution of the SCB operators $\Phi_J$ and $p_J$ as well as the out-field, in terms of the in-field. However, to solve these nonlinear equations straightforwardly, we need to make some approximations.


\subsection{Voltage biased SCB approximation}
\label{subsec:voltage_biased_SCB}
In the following, we will neglect the last term  in (\ref{fullPhiOutEq}). Since the time-derivative enters in product with $\tau_{RC}$, this will be a good approximation as long as the relevant frequencies of the incoming field $\Phi^{in}$ and of the SCB dynamics ($p_J$) is much lower than the inverse RC-time. 
Under this approximation, the final equations of motion are:
\begin{eqnarray}
\label{OurEquationOfMotion}
\partial_t \Phi_J &=& \frac{p_J+ C_c
  \left(V_{DC}+2 \partial_t \Phi^{in} +\frac{\tau_{RC}}{C_J}  \partial_t
    p_J \right)}{C_\Sigma}, \\
\label{OurEquationOfMotion2}
\partial_t p_J&=&-E_J \frac{2e}{\hbar}\sin{\left(\frac{2e}{\hbar}\Phi_J\right)}, \\
\label{OurOutField}
 \Phi^{out} &=& \Phi^{in} + \frac{\tau_{RC}}{C_J} p_J.
\end{eqnarray}
Here, we also note that this approximation is valid in recent experiments\cite{hoi11,hoi12}, where $Z_0=50 \Omega$, $C_c \sim 10$ fF and $C_J \sim 25$ fF, giving an inverse RC-timescale  of $1/(2\pi\tau_{RC}) \sim$ 400 GHz, which is around 50 times higher than the relevant frequency of $\Phi^{in}$ and $p_J$, set by the qubit frequency  $\sim 7.5$ GHz .

The above set of equations (\ref{OurEquationOfMotion}-\ref{OurOutField}) correspond to the Hamiltonian
\begin{eqnarray}
H &=& H_{sys}+H_{int}+H_{bath}, \label{OurFinalTotalHamiltonian}\\
H_{sys} &=& \frac{\left[p_J+C_c V_{DC}\right]^2}{2C_\Sigma}-E_J
\cos{\left(\frac{2e}{\hbar}\Phi_J\right)}, \label{SCBhamiltonian} \\
H_{int}&=& \frac{C_c}{C_\Sigma} (p_J+C_cV_{DC}) \partial_t \Phi(0^-,t),  \\
H_{bath}&=& \frac{\left[C_c \partial_t \Phi(0^-,t)\right]^2}{2C_\Sigma}+ \int_{-\infty}^0 \frac{p(x,t)^2}{2C_0} + \frac{\left[\partial_x \Phi(x,t)\right]^2}{2L_0} dx .
\end{eqnarray}
Thus we have arrived at the Hamiltonian of a voltage biased SCB, weakly coupled to the transmission line voltage at $x=0$, i.e. $V_0(t)=\partial_t \Phi_0(t)$. (Here, we note that for the uncoupled transmission line, without SCB, $\Phi_0(t)=2\Phi^{in}(t)$ due to the perfect reflection.) Truncating the Hilbert space of $H_{sys}$ to two levels, (\ref{OurFinalTotalHamiltonian}) is just the spin-boson Hamiltonian.
From this point we can proceed with a  Bloch-Redfield derivation of a master equation for the SCB only\cite{Rau2004}. By comparing to section 3.2. in \cite{gardinerzoller}, we also note that the equations of motion (\ref{OurEquationOfMotion}-\ref{OurEquationOfMotion2}) can be interpreted as quantum Langevin
equations (QLE) of the form
\begin{equation}
\dot{Y}=\frac{i}{\hbar}[H_{sys},Y]+\frac{i}{2\hbar}[\gamma
\dot{X}-2\sqrt{\gamma v}\dot{A}^{in},[X,Y]]_+,
\label{QLE}
\end{equation}
whereas (\ref{OurOutField}) stands for the input-output relation
\begin{equation}
A^{out}(t)=A^{in}(t)-\sqrt{\frac{\gamma}{v}}X(t),
\end{equation}
using the identifications $Y= \Phi_J$, $X= -(p_J+C_cV_{DC})$, $A^{in}=\sqrt{C_0} \Phi^{in}$ and where
\begin{equation}
\gamma=Z_0\left(\frac{C_c}{C_\Sigma}\right)^2
\label{damping}
\end{equation} 
is the damping constant that accounts for spontaneous emission.



\subsection{Master equation}
\label{subsection:MasterEquation}
From the QLE (\ref{QLE}) we can derive a master equation for the reduced density matrix of the SCB in the transmon regime. As input field, we first consider a thermal background at temperature $T$, giving rise to a photon occupation number of
\begin{equation} 
n_\omega=\frac{1}{\exp{\left(\hbar\omega/k_B T\right)}-1}.
\end{equation}
We assume that the density matrix initially can be written as a direct product, as well as Markovian properties and short correlation times for the transmission line variables. In the case when the damping (\ref{damping}) is much smaller than the system eigenenergies we arrive, after also employing the rotating wave approximation, at the following quantum optical master equation,
\begin{eqnarray}
\dot{\rho}(t)=&-&\frac{i}{\hbar} [ H_{sys},\rho ] \nonumber\\
&+&\frac{2\gamma}{\hbar}\sum_m \omega_m \left[\left(
      n_{\omega_m}+1
    \right) {\cal D}(X^-_m)\rho +  n_{\omega_m} {\cal D}(X^+_m)\rho\right],
\label{Master_Eq_RW}
\end{eqnarray}
with the Lindblad operator defined by $\mathcal{D}(c)\rho=c\rho c^{\dagger}-\frac{1}{2}\lr c^{\dagger}c\rho+\rho c^{\dagger}c\rr$. Also, $X$ has been decomposed into eigenoperators of $H_{sys}$,
\be
\ls H_{sys} , X_m^{\pm}\rs=\pm\hbar\omega_mX_m^{\pm},\hspace{5pt}\omega_m>0,
\ee
which is always possible as long as the eigenstates of $H_{sys}$ form a complete set.

Projecting the master equation onto the SCB eigenstates $|i\rangle$, $H_{\rm sys}|i\rangle=\omega_i|i\rangle$, ($i\in\{0,1,2,\dots\}$), we arrive at the following equation for the diagonal elements,
\begin{eqnarray}
\dot{\rho}_{ii}=&&\sum_{j\neq i} \Gamma_{ji} \rho_{jj} - \Gamma_{ij}
\rho_{ii}
\end{eqnarray}
where the relaxation ($\omega_{ij}=\omega_i-\omega_j >0 $ ) rates are 
\begin{equation}
\Gamma_{ij}= \frac{2\gamma}{\hbar} \omega_{ij} \left(1+
  n_{\omega_{ij}}\right) |\bra{i}X\ket{j}|^2,
\label{eq:relaxationrates}
\end{equation}
and the excitation ($\omega_{ij} < 0$) rates are
\begin{equation}
\Gamma_{ij}= \frac{2\gamma}{\hbar} |\omega_{ij}| n_{\omega_{ij}}
|\bra{i}X\ket{j}|^2.
\end{equation}

Noting that $X=-(p_J+C_cV_{DC})$ is the charge operator, the matrix elements can be calculated numerically from the SCB Hamiltonian in (\ref{SCBhamiltonian}). Denoting the SCB charging energy $E_C=e^2/2C_\Sigma$, the transmon regime is found for $E_J \gg E_C$ \cite{koch07}. Here, the SCB spectrum approaches a linear oscillator with the junction plasma frequency $\omega_p=\sqrt{8E_JE_C}/\hbar$ , and the charge operator asymptotically couples only neighboring eigenstates\cite{koch07}. We find the non-zero relaxation rates
\begin{equation}
 \Gamma_{(j+1)j}= \pi (j+1) \kappa^2  \frac{E_J}{\hbar}
 \frac{Z_0}{R_K} (1+n_{\omega_p}),
\end{equation}
and excitation rates
\begin{equation}
\Gamma_{j(j+1)}= \pi (j+1) \kappa^2  \frac{E_J}{\hbar} \frac{Z_0}{R_K}
n_{\omega_p},
\end{equation}
where $R_K=h/e^2 \approx 25$ k$\Omega$ denotes the quantum of resistance. The off-diagonal $ (i\neq j)$ elements are subject to a pure exponential decay, 
\be
\dot{\rho}_{ij}=-\gamma_{ij} \rho_{ij}
\end{equation}
with dephasing rates
\begin{equation}
\gamma_{ij}=\Gamma^{i}_\phi+\Gamma^{j}_\phi+\frac{1}{2}\left(\sum_{k\neq
    i} \Gamma_{ik} + \sum_{k\neq j} \Gamma_{jk}\right),
\label{eq:dephasingrates}
\end{equation}
equal to half the sum of all rates for transitions {\em from} state $|i\rangle$ and $|j\rangle$, as well as the pure dephasing rates
\begin{equation}
\Gamma^{k}_\phi=\frac{2\gamma}{\hbar}\frac{k_B T}{\hbar} |\langle
k|X|k\rangle|^2.
\end{equation}
The pure dephasing rates depend on the DC voltage, through the SCB spectrum, according to
\begin{equation}
|\langle k|X|k\rangle|=\frac{e}{4E_C}\left|\frac{\partial
    \omega_k(n_g)}{\partial n_g}\right|,
\end{equation}
where $n_g=C_c V_{DC}/2e$ is the dimensionless gate charge of the SCB.
In the transmon regime the spectrum is well approximated by
\begin{equation}
\omega_k(n_g)=\omega_k(n_g=1/4)-\frac{\epsilon_k}{2}\cos{(2\pi n_g)},
\end{equation}
where
\begin{equation}
\epsilon_k \simeq
(-1)^kE_C\frac{2^{4k+5}}{k!}\sqrt{\frac{2}{\pi}}\left
  (\frac{E_J}{2E_C}\right)^{\frac{k}{2}+\frac{3}{4}}e^{-\sqrt{8E_J/E_C}},
\end{equation}
giving a maximum thermal pure dephasing rate (for $n_g=\pm 1/4$) of
\begin{equation}
\max{\Gamma^{k}_\phi}=\kappa^2  \frac{Z_0}{R_K}  \frac{k_B T}{\hbar}
\frac{\pi^3}{8} \left|\frac{\epsilon_k}{E_C}\right|^2.
\end{equation}

Here, we also note that in addition to small amplitude thermal charge noise there can also be a slow but large amplitude charge drift. In some cases, the effect of this drift can be taken into account by averaging over the range of transition frequencies involved. In the transmon regime, for the transition from $|k\rangle$ to $|k+1\rangle$ this is given by $\epsilon_{k+1}-\epsilon_k\approx\epsilon_{k+1}$.

\subsection{Coherent drive}
\label{sec:coherentdrive}
In the next chapter, we will examine the scattering of coherent signals on the transmon in the two-level and three-level approximations. To include a coherent drive in the description, we take the input field $\Phi^{in}(t)$ to consist of a classical part $\Phi^{in}_{cl}(t)$ on top of the thermal background. Deriving the master equation for this case, it turns out that  (\ref{Master_Eq_RW}) is modified by adding the following time-dependent term to the system Hamiltonian,
\be
H_{d}(t)=-2\sqrt{\frac{\gamma}{Z_0}}\dot{\Phi}_{cl}^{in}(t)X.
\ee
\subsection{Adding more transmission lines}
\label{sec:morelines}
\begin{figure}[ht]
\centering
\subfigure[]{
	\includegraphics[width=0.8\columnwidth]{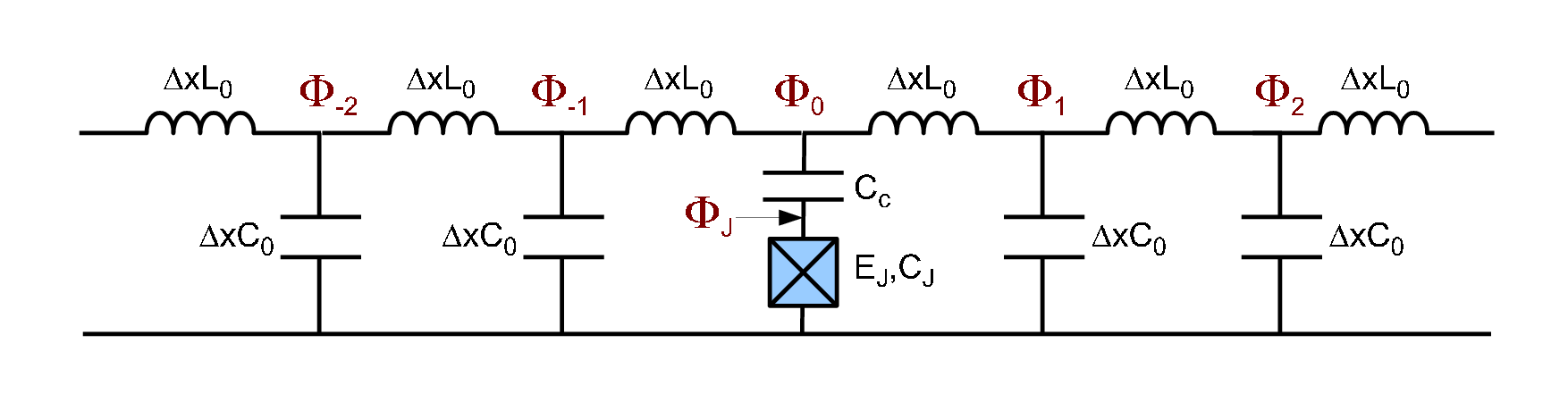}
	\label{fig:inf_circuit}
}
\subfigure[]{
	\includegraphics[width=\columnwidth]{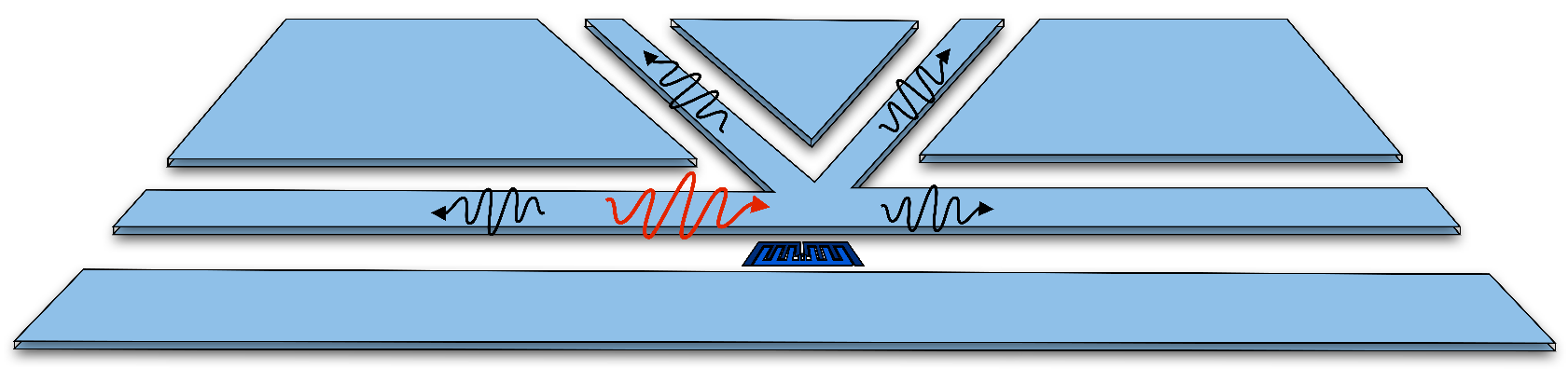}
	\label{fig:severalpaths}
}
\caption[]{(a) Discretized circuit describing the interaction
  of a single Cooper-pair box with microwave
  photons propagating in an infinite transmission line. (b) Generalization of the input-output formalism to an arbitrary number of ports connected by an artificial atom.}
\label{fig:subfigureExample}
\end{figure}

In this section, we generalize the above master equation by adding more semi-infinite transmission lines to the SCB.
First, by adding one more semi-infinite line, we arrive at the important case of an SCB capacitively coupled to an infinite transmission line. 
The discretized circuit is shown in Figure \ref{fig:inf_circuit}, and the corresponding Hamiltonian is obtained from (\ref{DiscreteHamiltonian}) by adding the transmission line terms for $x>0$
\be
H'_{d}=H_d + \frac{1}{\Delta x} \sum_{i > 0} \left( \frac{p_{i}^2}{2C_0} + \frac{(\Phi_{i-1} -\Phi_{i} )^2}{2L_0}\right).
\ee
From a similar analysis as above, we arrive at exactly the same master equation for the transmon's reduced density matrix, with the replacements
\begin{equation}
\Phi^{in}=\frac{1}{2}\left(\Phi^{in}_L+\Phi^{in}_R\right), \ \tau_{RC}=\frac{Z_0}{2}\frac{C_c C_J }{C_\Sigma},\ \gamma=\frac{Z_0}{2}\left(\frac{C_c}{C_\Sigma}\right)^2,
\label{eq:damping2}
\end{equation}
and the output fields are obtained from
\be
\Phi^{out}_{L/R}= \Phi^{in}_{R/L} + (\tau_{RC}/C_J) p_J.
\label{inoutLR}
\ee
We note that the damping constant $\gamma$ as well as the RC-time $\tau_{RC}$ are both halved compared to the semi-infinite case, since the impedance to ground is halved to $Z_0/2$. The in-field is the sum of the fields incoming from the left and right, but compared to the semi-infinite case the coupling coefficient is halved, since there is (almost) no reflection at $x=0$. Indeed, for a more general scenario with $N$ symmetrically coupled incident fields, as illustrated in Figure \ref{fig:severalpaths},
the mapping would be 
\begin{equation}
\Phi^{in}=\frac{1}{N}\sum_{n=1}^N\Phi^{in}_n,
\tau_{RC}=\frac{Z_0}{N}\frac{C_c C_J}{C_\Sigma},\ \gamma=\frac{Z_0}{N}\left(\frac{C_c}{C_\Sigma}\right)^2,
\end{equation}
and using the relation $\Phi_0=\Phi^{in}_n+\Phi^{out}_n\ (\forall n)$, the output fields are given as
\be
\Phi^{out}_n=\Phi_0-\Phi^{in}_n=\left(\frac{2}{N}-1\right)\Phi^{in}_n+\frac{\tau_{RC}}{C_J}p_J+\frac{2}{N} \sum_{m\neq n}^N \Phi^{in}_m.
\label{inoutn}
\ee

\section{Applications: scattering by the transmon}
\label{sec:applications}
\subsection{Two-level dynamics}
\label{sec:SCB}
In this section, we examine the scattering of coherent signals on the transmon in an open transmission line. The input field is a constant coherent signal with a single frequency $\omega_p$, close to resonance with the first transition frequency $\omega_{10}$ of the transmon. Thus, we can safely describe the transmon as a two-level system. The master equation is given by (\ref{Master_Eq_RW}) with a coherent drive and generalized to the case of an infinite transmission line (see sections  \ref{sec:coherentdrive} and \ref{sec:morelines}), in the special case of only one system eigenfrequency $\omega_{10}$. Moreover, we include an additional term due to pure dephasing, so that the total dephasing rates are given by (\ref{eq:dephasingrates}).
We represent our operators by the following Pauli matrices (using the notation $X_{ij}\equiv\bra{i}X\ket{j}$)
\be
 H_{sys}=-\hbar\frac{\omega_{10}}{2}\sigma_z, \hspace{5pt} X^{\pm}=\pm i|X_{10}|\sigma^{\pm}.
\ee
Below, we will determine reflection and transmission coefficients for coherent signals scattered on the transmon. In the previous section the incoming and outgoing fields were described in terms of the flux, since that gives a simpler description of the transmon. However, the voltage is a more intuitive quantity than the flux and it is also usually what is measured in experiments. Therefore, in this section, we will describe the inputs and outputs in terms of the voltage.

We consider an incoming coherent voltage field, 
\be
V^{in}_L(t)=\Omega_p\sin{\omega_pt},
\label{vinleft}
\ee
impinging on the transmon from the left. For simplicity, we set the temperature to zero ($n_{\omega_{10}}=0$). 
The reflected voltage field is the output to the left of the transmon. Using (\ref{eq:damping2}) and (\ref{inoutLR}), we have
\be
V^{out}_L(t)=-\sqrt{\frac{\gamma Z_0}{2}}\langle\dot{X}(t)\rangle,
\label{vreflected}
\ee
where the expectation value can be written as
\be
\langle \dot{X}(t)\rangle=i\omega_{10}\lr\langle X^+(t)\rangle-\langle X^-(t)\rangle\rr=-\omega_{10}|X_{10}|\langle\sigma^x\rangle.
\label{eq:xexp}
\ee
Inserting (\ref{eq:xexp}) into (\ref{vreflected}) yields
\be
V^{out}_L(t)=\half\sqrt{\hbar\omega_{10}\Gamma_{10}Z_0}\langle\sigma^x\rangle=\sqrt{\hbar\omega_{10}\Gamma_{10}Z_0}\textit{Re}\ls\rho_{01}\rs,
\label{eq:vreflected2}
\ee
where $\rho_{01}$ is a density matrix element in the transmon eigenbasis.

\begin{figure}[t]
\begin{center}\includegraphics[width=0.4\columnwidth]{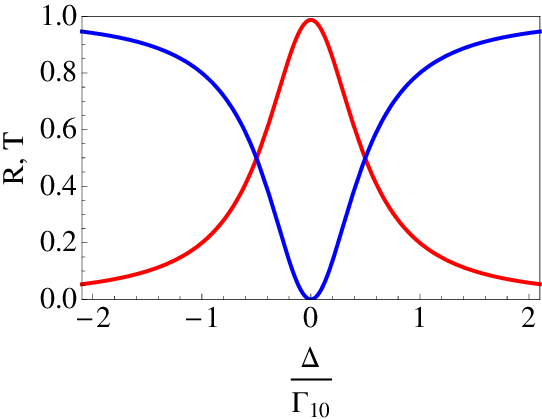}
\caption{Reflectance $R$ (red) and transmittance $T$ (blue) for a two-level transmon as a function of detuning, with the average number of incoming photons per interaction time being $N_{in}/(\Gamma_{10}/2\pi)=0.01$.} 
\label{fig:TwoLevelPlot}
\end{center}
\end{figure}

To solve the master equation we perform a unitary transformation to a frame rotating with the driving frequency  $\omega_p$. In this frame, the equation becomes time-independent after employing the rotating-wave approximation. Solving the equation in the steady-state ($\dot{\rho}=0$) and transforming back to the non-rotaing frame yields the following expression for the desired density matrix element,
\bea
\rho_{01}&=&\half\frac{\sqrt{\hbar\omega_{10}\Gamma_{10}Z_0}\lr \Delta+i\gamma_{10}\rr\Omega_p}{\hbar\omega_{10}Z_0\gamma_{10}^2+\hbar\omega_{10}Z_0\Delta^2+\gamma_{10}\Omega_p^2}e^{i\omega_pt},
\label{eq:rho01}
\eea
where $\Delta\equiv\omega_{p}-\omega_{10}$ is the detuning. Now, plugging this expression into (\ref{eq:vreflected2}) results in
\be
V_{L}^{out}(t)=-\frac{\Omega_p}{2}\frac{\sin{\omega_pt}-\frac{\Delta}{\gamma_{10}}\cos{\omega_pt}}{\frac{\gamma_{10}}{\Gamma_{10}}+\frac{\Delta^2}{\Gamma_{10}\gamma_{10}}+2\frac{N_{in},}{\Gamma_{10}}},
\ee
where $N_{in}=\Omega_p^2/(2Z_0\hbar\omega_{10})$ is the average number of incoming photons per second.
Thus, the reflection coefficient for the negative frequency part of the field is given by
\be
r=-r_0\frac{1-i\frac{\Delta}{\gamma_{10}}}{1+\lr\frac{\Delta}{\gamma_{10}}\rr^2+2\frac{N_{in}}{\gamma_{10}}},
\ee
with $r_0\equiv\Gamma_{10}/2\gamma_{10}$.
For the transmitted field, (\ref{inoutLR}) yields
\be
V^{out}_R(t)=V^{in}_L(t)-\sqrt{\frac{\gamma Z_0}{2}}\langle\dot{X}(t)\rangle,
\label{vtransmitted}
\ee
which directly gives us the following expression for the transmission coefficient,
\be
t=1+r=\frac{1-r_0+\lr\frac{\Delta}{\gamma_{10}}\rr^2+2\frac{N_{in}}{\gamma_{10}}+ir_0\frac{\Delta}{\gamma_{10}}}{1+\lr\frac{\Delta}{\gamma_{10}}\rr^2+2\frac{N_{in}}{\gamma_{10}}}.
\ee
In Figure \ref{fig:TwoLevelPlot} we plot the reflectance $R=|r|^2$ and transmittance $T=|t|^2$ as a function of the detuning, in the case of a weak input signal and no pure dephasing.
For a resonant drive ($\Delta=0$) we see that perfect reflection is approached, in agreement with \cite{shen05,shen05b,romero09}.


\subsection{Three-level dynamics}
\label{subsec:three-level}
In the previous section,  we showed that a low-amplitude input signal is totally reflected when it resonantly scatters off a transmon in the two-level approximation. In this section, we instead study the scattering off a transmon in the three-level approximation. By strongly driving the second transition, the transmon becomes transparent to frequencies in resonance with the first transition. This effect is due to Autler-Townes splitting and has been observed in recent experiments \cite{hoi11}.

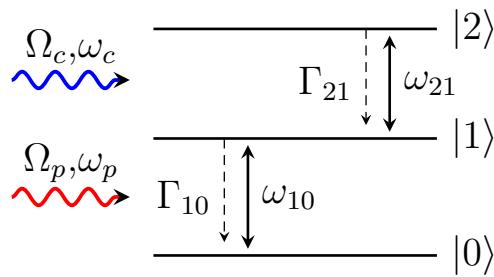
\begin{figure}[t]
\centerline{
  \resizebox{7cm}{!}{
    \begin{tikzpicture}[
      scale=0.5,
      level/.style={thick},
      virtual/.style={thick,densely dashed},
      trans/.style={thick,<->,shorten >=2pt,shorten <=2pt,>=stealth},
      classical/.style={thin,densely dashed,->,shorten >=4pt,>=stealth},
photon/.style={->,decorate, decoration={snake}, draw=blue,
      line width=0.4mm,>=stealth},
photon2/.style={->,decorate, decoration={snake}, draw=red,
      line width=0.4mm,>=stealth},
]
    \draw[level] (-3cm,0em) -- (3cm,0em) node[right] {$\ket{0}$};
    \draw[level] (-3cm,6em) -- (3cm,6em) node[right] {$\ket{1}$};
    \draw[level] (-3cm,11.62em) -- (3cm,11.62em) node[right] {$\ket{2}$};
    \draw[trans] (2cm,11.62em) -- (2cm,6em) node[midway,right] {$\omega_{21}$};
    \draw[trans] (-1cm,6em) -- (-1cm,0em) node[midway,right] {$\omega_{10}$};
    \draw[classical] (1.5cm,11.62em) -- (1.5cm,6em) node[midway,left] {$\Gamma_{21}$};
 \draw[classical] (-1.5cm,6em) -- (-1.5cm,0em) node[midway,left] {$\Gamma_{10}$};
 \draw[photon] (-6cm,9em) -- (-3.5cm,9em) node[midway,above] {$\Omega_c$,$\omega_c$};
 \draw[photon2] (-6cm,3em) -- (-3.5cm,3em) node[midway,above] {$\Omega_p$,$\omega_p$};
    \end{tikzpicture}
  }
}
\caption{Internal levels of the transmon in the three-level approximation. A strong control field drives the $\ket{1}\rightarrow\ket{2}$ transition, rendering a transparency for the $\ket{0}\rightarrow\ket{1}$ transition.}
\label{leveldiagram}
\end{figure}

We consider an incoming voltage field from the left, consisting of a probe field $\Omega_p\sin{\omega_pt}$ close to resonance with the first transition (with detuning $\Delta_p=\omega_{p}-\omega_{10}$) and a control field $\Omega_c\sin{\omega_ct}$ close to resonance with the second transition (with detuning $\Delta_c=\omega_{c}-\omega_{21}$). Figure \ref{leveldiagram} shows the energy  levels of the transmon in this approximation.
In the transmon eigenbasis, the relevant operators are (with the ground state energy $\omega_0=0$)
\begin{eqnarray}
H_{sys}&=&\hbar\sum_{i=1}^2\omega_i \ket{i}\bra{i},\\
X&=&i\sum_{i=1}^{2}|X_{i(i-1)}|(\sigma_i^+-\sigma_i^-),
\end{eqnarray}
with $\sigma^+_i=\ket{i}\bra{i-1}$ and $\sigma^-_i=(\sigma_i^+)^{\dagger}$. 
In the same way as in the two-level case ((\ref{eq:xexp})-(\ref{eq:vreflected2})), we obtain the following expression for the reflected signal
\bea
V^{out}_L(t)=-\sqrt{\frac{\gamma Z_0}{2}}\langle\dot{X}(t)\rangle=\half\sum_{i=1}^2\sqrt{\hbar\omega_{i(i-1)}Z_0\Gamma_{i(i-1)}}\langle\sigma_i^x\rangle,
\eea
with $\sigma_i^x=\sigma_i^++\sigma_i^-$.
Thus, the reflected field consists of one part with frequencies around the the probe frequency $\omega_p$ and one part with frequencies around the control frequency $\omega_c$. Since we are interested in the reflectance and transmittance properties of the probe, we concentrate on the corresponding part of the reflected field,
\bea
V_{p}^{ref}(t)=\half\sqrt{\hbar\omega_{10}Z_0\Gamma_{10}}\langle\sigma_1^x\rangle=\sqrt{\hbar\omega_{10}Z_0\Gamma_{10}}\textit{Re}(\rho_{10}).
\label{Vrefprobe}
\eea
The master equation is given by (\ref{Master_Eq_RW}) for the case of two system eigenfrequencies, again with a coherent drive and generalized to the case of an infinite transmission line (see sections \ref{sec:coherentdrive} and \ref{sec:morelines}). Also, terms accounting for pure dephasing are added. To transform the master equation to a time-independent picture, we use the following unitary transformation matrix,
\begin{equation}
U(t)= \left( \begin{array}{ccc}
    1& 0 & 0 \\
0 & e^{-i\omega_p t}& 0 \\
0 & 0 & e^{-i(\omega_p+\omega_c)t} \end{array} \right),
\end{equation}
and employ the rotating-wave approximation. As before, we solve the master equation in the steady-state to determine $\rho_{10}$, but we now consider two different cases. 

Firstly, by setting $\Omega_c=0$, we recover exactly the same expression for the reflected field as in the two-level case. Thus, with the control field turned off, we see almost full reflection for weak probe fields on resonance with the first transition frequency of the transmon. 

Secondly, we consider the case of a strong control field ($\Omega_c\gg\Omega_p$). Solving the master equation and expanding $\rho_{10}$ to first order in $(\Omega_p/\Omega_c)$, we obtain the following expression,
\begin{figure}[ht]
\centering
\subfigure[]{
	\includegraphics[width=0.4\columnwidth]{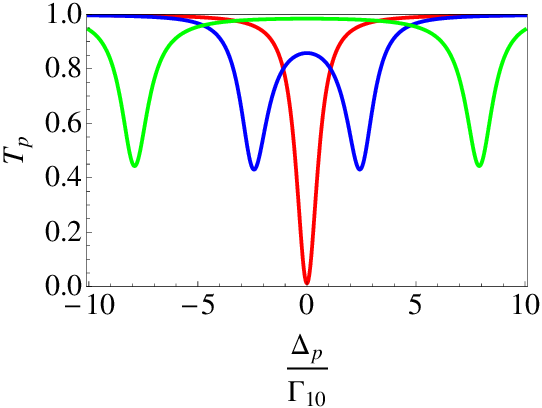}
	\label{fig:T_EIT1}
}
\subfigure[]{
	\includegraphics[width=0.4\columnwidth]{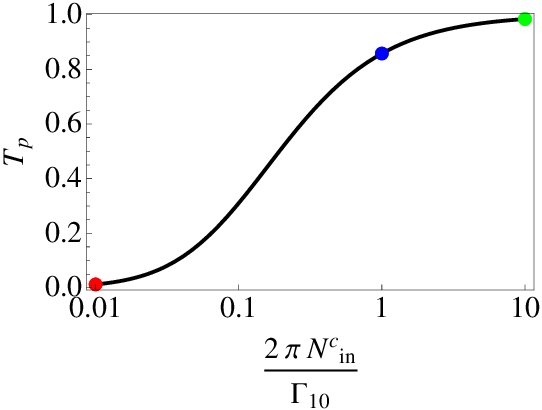}
	\label{fig:T_EIT2}
}
\caption[]{(a) Transmittance $T_p$ of the probe as a function of detuning for three different control field strengths; $N_{in}^c/(\Gamma_{10}/(2\pi))=0.01$ (red), $N_{in}^c/(\Gamma_{10}/(2\pi))=1$ (blue) and $N_{in}^c/(\Gamma_{10}/(2\pi))=8$ (green). (b) Transmittance as a function of control field strength for a resonant probe field $(\Delta_p=0)$.}
\label{fig:T_EIT}
\end{figure}
\be
\rho_{10}^{(1)}=-\frac{2i\hbar\omega_{21}Z_0\sqrt{\frac{\Gamma_{10}}{\hbar\omega_{10}Z_0}}\lr\gamma_{20}-i\lr\Delta_c+\Delta_p\rr\rr\Omega_p}{4\hbar\omega_{21}Z_0\lr\gamma_{10}-i\Delta_p\rr\lr\gamma_{20}-i\lr\Delta_c+\Delta_p\rr\rr+\Gamma_{21}\Omega_c^2}e^{-i\omega_pt}.
\label{rho10firstorder}
\ee
Inserting (\ref{rho10firstorder}) into (\ref{Vrefprobe}) we can determine the reflection coefficient. For a resonant control field ($\Delta_c=0$), the result is
\be
r=-\frac{2\Gamma_{10}\lr\gamma_{20}^2+\Delta_p^2\rr\lr\gamma_{10}-i\Delta_p\rr+\Gamma_{10}\Gamma_{21}\lr\gamma_{20}+i\Delta_p\rr N^c_{in}}{4\lr\gamma_{10}^2+\Delta_p^2\rr\lr\gamma_{20}^2+\Delta_p^2\rr+4\Gamma_{21}\lr\gamma_{10}\gamma_{20}-\Delta_p^2\rr N_{in}^c+\Gamma_{21}^2N_{in}^{c2}}
\ee
where $N^c_{in}=\Omega_c^2/(2Z_0\hbar\omega_{21})$ is the average number of incoming photons per second in the control field. The transmission coefficient is again given by $t=1+r$. Figure \ref{fig:T_EIT} shows the transmittance $T=|t|^2$ for different probe detunings and control field strengths. In these plots, we have neglected pure dephasing and used (\ref{eq:relaxationrates}) to express $\Gamma_{21}$ in terms of $\Gamma_{10}$.

We can clearly see that, for strong control fields, the transmittance of a resonant probe approaches unity. Thus, by turning on and off a strong resonant control field, we can switch between the cases of full transmission and full reflection for the resonant probe.

\subsection{Second-order correlations}
\label{sec:second-order}
In a recent experiment \cite{hoi12}, the second-order statistics of the field scattered off a transmon was measured. In this section, inspired by the experiment, we analyze the second-order correlation functions in our system.
\begin{figure}[!ht]
  \begin{center}
 \includegraphics[width=0.7 \columnwidth]{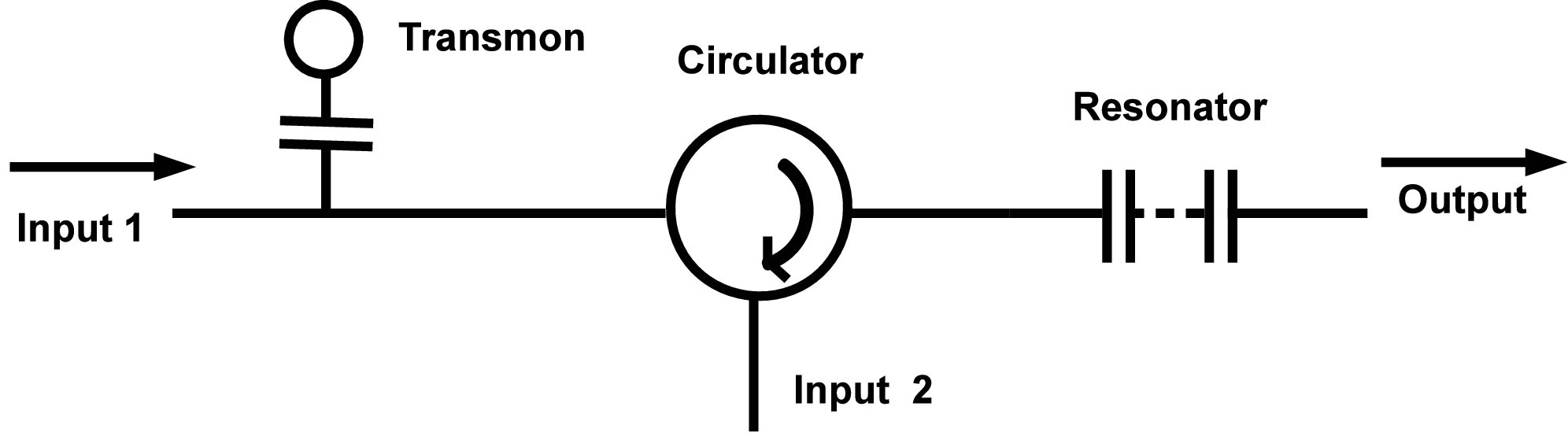}
\caption{Schematic model of a transmon cascaded with a resonator. The circulator prevents the field
reflected from the resonator to reach the transmon. Using input port 1(2), the output is the filtered transmitted (reflected) signal.} 
\label{fig:cascading}
\end{center}
\end{figure}
The normalized second-order correlation function is in the steady state given by \cite{gerryknight}
\be
g^{(2)}(\tau)=\frac{\langle V^+(t)V^+(t+\tau)V^-(t+\tau)V^-(t)\rangle}{\langle V^+(t)V^-(t)\rangle^2},
\ee
and is proportional to the conditional probability of detecting a photon at time $t+\tau$, given that one was detected at time $t$. Here,  $V^{\pm}(t)$ are the positive and negative frequency parts of the voltage field.

We calculate $g^{(2)}(\tau)$ for the transmitted and reflected fields from a transmon driven by a resonant coherent signal. We treat the transmon as a two-level system and use the same notation as in section \ref{sec:SCB}. To be able to compare with the experiments in \cite{hoi12} we perform the calculations for finite temperatures and a finite detection bandwidth on the output signal. For zero temperature and infinite bandwidth we recover the results of \cite{lukin07}; perfect antibunching in the reflected field and bunching in the transmitted field.

Including the effect of a finite detection bandwidth is straighforward, by including a filter in the calculations. The approach we have taken is to model the filter by a single-mode transmission line resonator in resonance with the transmon, with the Hamiltonian
\be
H_{res}=\hbar\omega_{10}a^{\dagger}a,
\ee
and cascade it with the transmon. 
 We start from the quantum Langevin equation (\ref{QLE}) for the transmon, generalized to the case of an infinite transmission line (see section \ref{sec:morelines}), and a similar equation for the resonator. Our coherent input signal is the voltage field $V^{in}_L(t)=\Omega_d\sin{\omega_d t}$, just like in section \ref{sec:SCB}. We can then use the formalism of cascaded quantum systems in \cite{gardinerzoller} to arrive at a master equation for the joint density matrix of the transmon and the resonator. In this formalism, the output from the transmon (reflected or transmitted) is taken as input to the resonator, without any signals going the opposite way (see Figure \ref{fig:cascading}). For the field reflected from the transmon, the resulting master equation is
\bea
\dot{\rho}&=&\frac{i}{\hbar}[\rho,H_{sys}+H_{res}]+\Gamma_{10}\mathcal{D}(\sigma^-)\rho+\Gamma_{01}\mathcal{D}(\sigma^+)\rho\nonumber\\
&&+\gamma_{BW}\ls\lr\frac{n_{\omega_{10}}}{2}+1\rr\mathcal{D}(a)\rho+\frac{n_{\omega_{10}}}{2}\mathcal{D}(a^{\dagger})\rho\rs\nonumber\\
&&+\half i\sqrt{\Gamma_{10}(n_{\omega_{10}}+1)\gamma_{BW}}\lr [a,\rho \sigma^+]+[a^{\dagger},\sigma^-\rho]\rr\nonumber\\
&&+\half i\sqrt{\Gamma_{01}n_{\omega_{10}}\gamma_{BW}}\lr [\sigma^+\rho,a]+[\rho\sigma^-,a^{\dagger}]\rr\nonumber\\
&&+i\sqrt{\frac{\Gamma_{10}N_{in}}{2(n_{\omega_{10}}+1)}}[\rho,\sigma^x],
\label{cascadedmaster}
\eea
where we have denoted the filter bandwidth by $\gamma_{BW}$. For the field transmitted through the transmon, (\ref{cascadedmaster}) is modified by simply adding the following term to the right-hand side,
\be
\sqrt{\frac{\gamma_{BW}N_{in}}{2}}[\rho,a^{\dagger}-a].
\label{eq:transmasterterm}
\ee
The output we are interested in is the voltage field leaking out at the right side of the resonator, whose positive and negative frequency parts are proportional to $a(t)$ and $a^{\dagger}(t)$ respectively. Thus, $g^{(2)}(\tau)$ can be calculated as
\bea
g^{(2)}(\tau)&=&\frac{\langle a^{\dagger}(t)a^{\dagger}(t+\tau)a(t+\tau)a(t)\rangle}{\langle a^{\dagger}(t)a(t)\rangle^2}
=\frac{\Tr\ls a^{\dagger}aP(\tau)(a\rho_s a^{\dagger})\rs}{\Tr\ls a^{\dagger}a\rho_s\rs^2},
\label{g2cascaded}
\eea
where $\rho_s$ is the steady-state density matrix and $P(\tau)$ the propagator super-operator, defined by $\rho(t+\tau)=P(\tau)\rho(t)$. Both $\rho_s$ and $P(\tau)$ are obtained by solving the master equation (\ref{cascadedmaster})-(\ref{eq:transmasterterm}). For the case without filter, $g^{(2)}(\tau)$ for the reflected field is given by (\ref{g2cascaded}) with $a$ replaced by $\sigma^-$. Since $(\sigma^-)^2=0$, it directly follows that $g^{(2)}(0)=0$, \textit{i.e.} perfect antibunching.

\begin{figure}[ht]
\centering
\subfigure[]{
	\includegraphics[width=0.4\columnwidth]{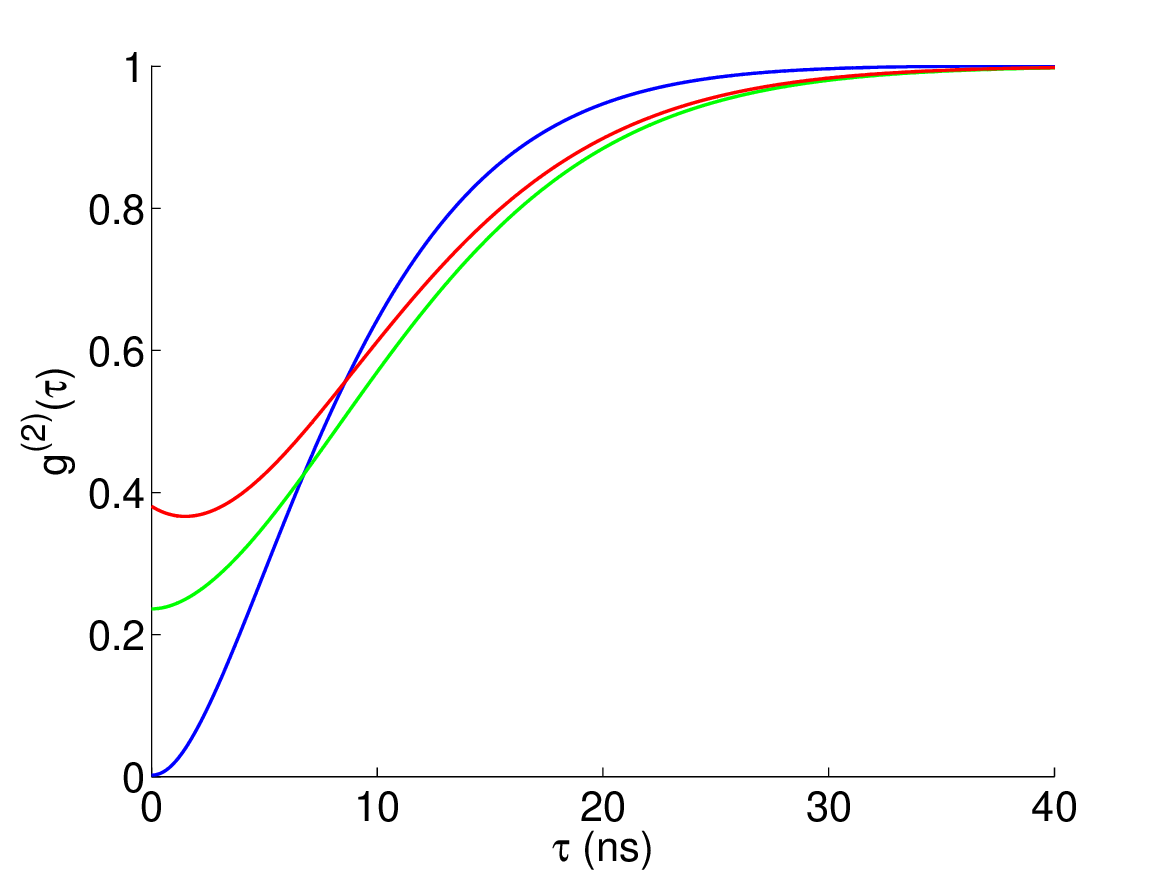}
	\label{fig:antibunching}
}
\subfigure[]{
	\includegraphics[width=0.4\columnwidth]{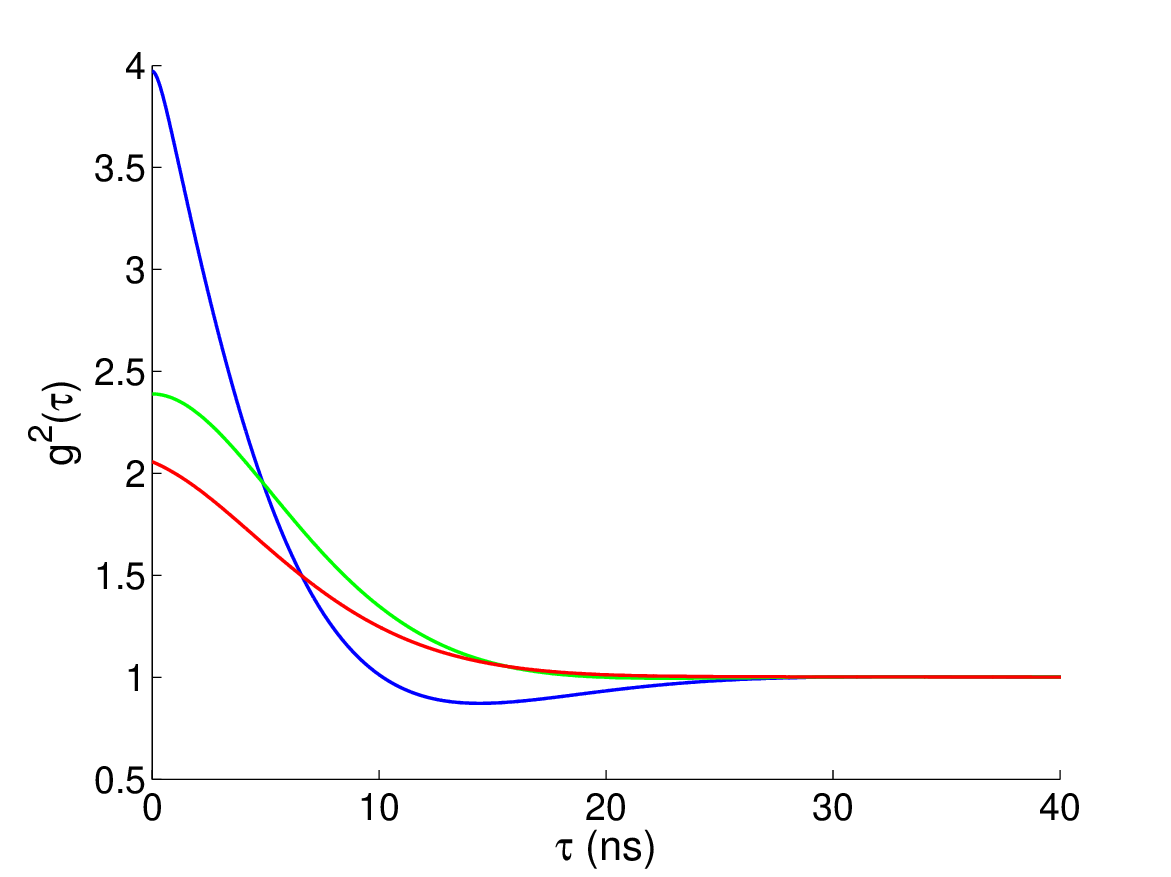}
	\label{fig:bunching}
}
\caption[]{$g^{(2)}(\tau)$ for the fields reflected from and transmitted through a transmon for different temperatures and detection bandwidths. Typical parameter values from recent experiments are used ($\Gamma_{10}/2\pi=41$ MHz, $\omega_{10}/2\pi=5.12$ GHz). (a) Reflected field: blue ($T=0$, BW$=1$ GHz), green ($T=0$, BW$=55$ MHz), red ($T=50$ mK, BW$=55$ MHz), with P=-131 dBm. (b) Transmitted field: blue ($T=0$, BW$=1$ GHz), green ($T=0$, BW$=55$ MHz), red ($T=80$ mK, BW$=55$ MHz), with P=-127 dBm. }
\label{fig:g2}
\end{figure}

In Figure \ref{fig:antibunching} we plot $g^{(2)}(\tau)$ for the reflected field for different temperatures and detection bandwidths. Typical parameter values from recent experiments \cite{hoi12} are used. For zero temperature and large bandwidth we see perfect antibunching, as expected. For a decreased bandwidth the full time dynamics of the antibunching cannot be resolved, which results in a less pronounced antibunching dip. For finite temperatures we see even less antibunching, due to a nonzero probability of detecting bunched thermal photons.
In Figure \ref{fig:bunching} we plot $g^{(2)}(\tau)$ for the transmitted field for different temperatures and detection bandwidths. Here, we see a decrease of the superbunching for higher temperatures and smaller bandwidths.
These results explain the qualitative features of the experimental data in \cite{hoi12} well. 
\section{Summary and conclusions}
\label{sec:conclusions}
Summing up, we have performed a thorough analysis of the qubit-photon
scattering in a one dimensional continuum from a microscopic point
of view.  In particular, we have derived a master equation description using a superconducting transmon qubit as our scatterer. When we consider the two lowest levels of the transmon it behaves as a mirror for the
incoming photons. Then, going beyond to the two-level
approximation, we can use a control field resonant with a second transition of the
transmon to suppress this reflection of photons at the probe frequency. 
Finally we discussed how the photon antibunching observed in the reflected field is reduced by finite temperature and finite detection bandwidth.

\section*{Acknowledgements}

We would like to thank T. Willemen and Tauno Palomaki for valuable discussions.
We acknowledge financial support from the Swedish Research Council, 
the Wallenberg foundation, STINT and from the EU through the ERC and the 
projects SOLID and PROMISCE.This work was also
supported by Spanish MICINN Project  FIS2009-10061, and CAM research
consortium QUITEMAD S2009-ESP-1594. 
B.P. acknowledges support from CSIC JAE-PREDOC2009 Grant. 
\\
\bibliographystyle{unsrt}


\bibliography{cqed}
\end{document}